\documentclass[useAMS,usenatbib]{mn2e}

\usepackage{epsfig}
\usepackage{amsmath}
\usepackage{amssymb}
\usepackage{natbib}
\usepackage{threeparttable} 

\newcommand{\Msun}{\mathrm{M}_{\odot}}
\newcommand{\lum}{\mathrm{erg~s}^{-1}}
\newcommand{\halfa}{$\mathrm{H\alpha}$}
\newcommand{\flux}{\mathrm{erg~cm}^{-2}~\mathrm{s}^{-1}}
\newcommand{\fluence}{\mathrm{erg~cm}^{-2}}
\newcommand{\cnts}{\mathrm{counts~s}^{-1}}
\newcommand{\kms}{\mathrm{km~s}^{-1}}

\newcommand{\rxh}{J1735}
\newcommand{\rxhname}{1RXH~J173523.7--354013}
\newcommand{\rx}{RX~J1735.3--3540}
\newcommand{\rxs}{1RXS~J173524.4--353957}
\newcommand{\igr}{IGR~J17353--3539}
\newcommand{\xte}{XTE~J1701--407}
\newcommand{\ucxb}{4U~1246--58}

\newcommand{\chan}{\textit{Chandra}}

\newcommand{\swift}{\textit{Swift}}
\newcommand{\xmm}{\textit{XMM-Newton}}
\newcommand{\rosat}{\textit{ROSAT}}
\newcommand{\inte}{\textit{INTEGRAL}}
\newcommand{\vlt}{\textit{VLT}}
\newcommand{\ntt}{\textit{NTT}}
\newcommand{\rem}{\textit{REM}}
\newcommand{\magellan}{\textit{Magellan}}

\def \mnras {MNRAS}
\def \apj {ApJ}
\def \apjs {ApJS}
\def \apjl {ApJL}
\def \aap {A\&A}

\def \pasp {PASP}
\def \aaps {AAPS}

\def\farcs{\hbox{$.\!\!^{\prime\prime}$}}

\title[\rxhname]{Multi-wavelength observations of \rxhname: revealing an unusual bursting neutron star
}
\author[N. Degenaar et al.]
{N. Degenaar$^{1}$\thanks{e-mail: degenaar@uva.nl}, P.G. Jonker$^{2,3}$, M.A.P. Torres$^{3}$, R.Kaur$^1$, N. Rea$^{1,4}$, G.L. Israel$^{5}$,
\newauthor A. Patruno$^{1}$, G. Trap$^{6,7}$, E.M. Cackett$^{8}$, P. D'Avanzo$^{9}$, G. Lo Curto$^{9}$, 
\newauthor G. Novara$^{10,11}$, H. Krimm$^{12,13}$, S.T. Holland$^{12,13}$, A. De Luca$^{10}$, P. Esposito$^{10}$, 
\newauthor R. Wijnands$^{1}$
\vspace{0.2cm}\\
$^{1}$Astronomical Institute ``Anton Pannekoek", University of Amsterdam, Postbus 94249, 1090 GE, Amsterdam, the Netherlands\\
$^{2}$SRON, Netherlands Institute for Space Research, Sorbonnelaan 2, 3584 CA, Utrecht, the Netherlands\\
$^{3}$Harvard-Smithsonian  Center for Astrophysics, 
60 Garden Street, 
Cambridge, MA~02138, U.S.A.\\
$^{4}$Institut de Ciencies de l'Espai (ICE-CSIC, IEEC), Campus UAB, Facultat de Ciencies, Torre C5-parell, 2a planta, 08193,\\
$^{~~}$Barcelona, Spain\\
$^{5}$INAF-Osservatorio Astronomico di Roma, Via Frascati 33, I-00040 Monteporzio Catone (Roma), Italy\\
$^{6}$Service d'Astrophysique (SAp) / IRFU / DSM / CEA Saclay, B\^{a}t. 709, 91191 Gif-sur-Yvette Cedex, France\\
$^{7}$AstroParticule \& Cosmologie (APC) / Universit\'e Paris VII / CNRS / CEA / Observatoire de Paris -- B\^{a}t. Condorcet, 10, \\
$^{~~}$rue Alice Domon et L\'eonie Duquet, 75205 Paris Cedex 13, France\\
$^{8}$\chan\ fellow, University of Michigan, Department of Astronomy, 500 Church St, Ann Arbor, MI 48109, USA\\
$^{9}$INAF-Osservatorio Astronomico di Brera, via Emilio Bianchi 46, 23807 Merate (LC), Italy\\
$^{10}$INAF-Istituto di Astrofisica Spaziale e Fisica Cosmica Milano, via Bassini 15, I-20133, Milano, Italy\\
$^{11}$Universit\`{a} degli Studi di Pavia, Dipartimento di Fisica Nucleare e Teorica, via Bassi 6, I-27100, Pavia, Italy\\
$^{12}$Universities Space Research Association, 
10211 Wincopin Circle, Suite 500, 
Columbia, MD 21044, USA\\
$^{13}$NASA/Goddard Space Flight Center, 8800 Greenbelt Road, Greenbelt, MD 20771, USA
}

\begin{document}

\date{Accepted 2010 January 19.  Received 2010 January 19; in original form 2009 December 2}

\pagerange{\pageref{firstpage}--\pageref{lastpage}} \pubyear{2009}

\maketitle

\label{firstpage}

\begin{abstract}
On 2008 May 14, the Burst Alert Telescope aboard the \swift\ mission triggered on a type-I X-ray burst from the previously unclassified \textit{ROSAT} object \rxhname, establishing the source as a neutron star X-ray binary. We report on X-ray, optical and near-infrared observations of this system. The X-ray burst had a duration of $\sim 2$~h and belongs to the class of rare, intermediately long type-I X-ray bursts. From the bolometric peak flux of $\sim 3.5 \times 10^{-8}~\flux$, we infer a source distance of $D\lesssim 9.5$~kpc. Photometry of the field reveals an optical counterpart that declined from $R = 15.9$ during the X-ray burst to $R = 18.9$ thereafter. Analysis of post-burst \swift/XRT observations, as well as archival \xmm\ and \rosat\ data suggests that the system is persistent at a  0.5--10 keV luminosity of $\sim2 \times 10^{35}~(D/\mathrm{9.5~kpc})^2~\lum$. Optical and infrared photometry together with the detection of a narrow \halfa\ emission line (FWHM$=292 \pm 9~\kms$, EW=$-9.0\pm0.4$~\AA) in the optical spectrum confirms that \rxhname\ is a neutron star low-mass X-ray binary. The \halfa\ emission demonstrates that the donor star is hydrogen-rich, which effectively rules out that this system is an ultra-compact X-ray binary.
\end{abstract}


\section{Introduction}\label{sec:intro}
The brightest Galactic X-ray point sources are X-ray binaries, in which either a neutron star or a black hole accretes mass from a companion star. When the accretion flow is continuous and the X-ray luminosity remains constant within a factor of a few, a system is classified as persistent. Transient X-ray binaries, on the other hand, alternate accretion outbursts that typically last for weeks to months with years to decades long episodes of quiescence, during which the X-ray luminosity is more than 2 orders of magnitude lower. 

One of the phenomena that uniquely mark the compact primary as a neutron star are type-I X-ray bursts (or shortly `X-ray bursts'); bright flashes of X-ray emission that are caused by unstable nuclear burning on the surface of the neutron star. They are characterized by blackbody emission with a peak temperature $kT_{\mathrm{bb}}>2$~keV and generally display a fast rise time followed by a slower decay phase. The initial rise can be interpreted as burning of the fuel layer, while the subsequent decay represents the cooling of the ashes. So far, X-ray bursts have only been detected from low-mass X-ray binaries (LMXBs), in which the donor star has a mass $M \lesssim 1~\Msun$.
The properties (e.g., duration, radiated energy and recurrence time) of type-I X-ray bursts depend on the conditions of the ignition layer, such as the temperature, thickness and hydrogen (H) abundance. These can drastically change as the mass-accretion rate onto the neutron star varies, which results in X-ray bursts with different characteristics for different accretion regimes \citep[for reviews, see e.g., ][]{lewin95,strohmayer06}. 

X-ray bursts can be serendipitously detected by the Burst Alert Telescope \citep[BAT;][]{barthelmy05} aboard the \swift\ satellite; a multi-wavelength observatory that is dedicated to the study of gamma-ray bursts (GRBs). 
Although events from known X-ray burst sources are ignored, the BAT occasionally triggers on an X-ray burst from a previously unknown burster \citep[e.g.,][]{zand08,linares09,wijnands09}.
On 2008 May 14 at 10:32:37 \textsc{UT}, \swift's BAT registered an X-ray flare \citep[][]{krimm08}. The BAT lightcurve and soft X-ray spectrum (no photons detected above $\sim35$ keV) suggested that this event was not a GRB \citep{baumgartner08,krimm08}. Rapid follow-up observations with the X-ray Telescope \citep[XRT;][]{burrows05} detected a bright, but quickly fading X-ray source within the 3 arcmin BAT error circle \citep{krimm08,baumgartner08}. Simultaneously obtained UV/Optical Telescope \citep[UVOT;][]{roming05} images revealed a fading optical source within the XRT error circle \citep{israel08}. 

The UVOT detection allowed for an accurate localization of the source of the BAT trigger:  $\alpha= \mathrm{17^{h}35^{m}23.75^{s}}$, $\delta=-35^{\circ} 40' 16.1\arcsec$ (J2000) with a 90 per cent confidence radius of 0.56 arcsec \citep{israel08}. Both the XRT and the UVOT position coincide with that of the unclassified X-ray source \rxhname\ (=\rxs=\rx; `\rxh' hereafter), which was discovered with the \rosat\ satellite in 1990.
The BAT trigger was likely the result of an X-ray burst from this system \citep{israel08}, and would thereby identify \rxh\ as a neutron star in, most likely, an LMXB. We note that \citet{rodriguez09} used \swift/XRT observations discussed in this paper to obtain a 3.5 arcsec position for the likely hard X-ray counterpart of \rxh, \igr\ (see Section~\ref{subsec:integral}). Based on that position, the authors identify a bright counterpart candidate in 2MASS (\textit{Ks}=$8.63\pm0.03$) and USNO-B1.0 (\textit{V}=$11.9\pm0.3$) catalogues, suggesting a possible high-mass X-ray binary nature. This object is also visible in our optical and near-infrared (near-IR) observations, but although it is very close to \rxh\ ($\sim4$~arcsec NW; see Fig.~\ref{fig:images}), it lies well outside the sub-arcsecond
UVOT position and is therefore not its counterpart.

In this paper we report on a multi-wavelength observing campaign of \rxh\ following the BAT trigger of 2008 May 14. We discuss the properties of the X-ray burst and the characteristics of the persistent emission. Our study comprises \swift\ data obtained with the BAT, XRT and UVOT, optical photometric observations carried out with the \textit{Rapid Eye Mount} (\rem) and the \textit{New Technology Telescope} (\ntt), optical spectroscopy using the \textit{Very Large Telescope} (\vlt), as well as near-IR observations performed with the \textit{Magellan Baade} telescope.  In addition, we explore archival \rosat, \inte\ and \xmm\ data to investigate the long-term flux and X-ray burst behavior of \rxh.

\section{Observations and data reduction}\label{sec:obs}
The observations that we obtained of \rxh\ with different facilities are listed in Table~\ref{tab:obs}. In the following sections these are discussed in more detail.

\begin{table*}
\begin{center}
\caption[l]{Observation log.}
\begin{threeparttable}
\begin{tabular}{l l l l l}
\hline \hline
Mission/Instrument & Observation ID & Date & Exposure time & Band\\
& & (UT) &  (ks) \\
\hline
\swift/BAT & 311603000 & 2008-05-14 &  0.5 & 15--150 keV\\
\textit{REM}/ROSS  & & 2008-05-14 & 0.15 & \textit{R} \\
\swift/XRT (WT) & 311603000 & 2008-05-14 &  $9.0\times 10^{-2}$ & 0.5--10 keV \\
\swift/UVOT & 311603000 & 2008-05-14 &  $7.7\times 10^{-2}$ & \textit{WH} ($\sim1500-8500$~\AA) \\
\swift/XRT (PC) & 311603000 & 2008-05-14 & 2.1 & 0.5--10 keV \\
\swift/UVOT  & 311603000 & 2008-05-14 &  0.39 & \textit{WH} ($\sim1500-8500$~\AA) \\
\swift/XRT (WT) & 311603001 & 2008-05-15 & 2.0 & 0.5--10 keV  \\
\swift/XRT (WT) & 311603002 & 2008-05-15 & 2.0 & 0.5--10 keV  \\
\textit{REM}/ROSS & & 2008-05-15 & 0.24 & \textit{R} \\
\textit{Magellan}/PANIC  & & 2008-05-25 & 1.2 & \textit{J, H, K} \\
\swift/XRT (PC) & 311603004 & 2008-05-28 & 4.6 & 0.5--10 keV  \\
\swift/XRT (PC) & 311603006 & 2008-06-05 & 4.4 & 0.5--10 keV  \\
\swift/XRT (PC)  & 311603008 &  2008-06-14 & 3.9 & 0.5--10 keV  \\
\textit{NTT}/EFOSC  & &  2008-06-16 & 3.9 & \textit{B, V, R}\\
\swift/XRT (PC)  & 311603009 &  2008-07-12 & 8.8 & 0.5--10 keV  \\
\vlt/FORS2 & & 2008-07-26/27 & 3.6 & $5300-8600$ \AA \\
\swift/XRT (PC)  & 311603011 &  2008-07-28 & 1.7 & 0.5--10 keV  \\
\swift/XRT (PC)  & 311603012 &  2008-07-29 & 2.2 & 0.5--10 keV  \\
\swift/XRT (PC)  & 311603013 &  2008-07-31 & 0.7 & 0.5--10 keV  \\
\swift/XRT (PC)  & 311603014 &  2008-08-02 & 1.9 & 0.5--10 keV  \\
\swift/XRT (PC)  & 311603015 &  2008-08-05 & 2.1 & 0.5--10 keV  \\
\swift/XRT (PC)  & 31446001 &  2009-07-24 &  1.8 & 0.5--10 keV \\
\hline
\end{tabular}
\label{tab:obs}
\end{threeparttable}
\end{center}
\end{table*}

\subsection{\swift}\label{subsec:swift}

\subsubsection{BAT}
\begin{figure}
 \begin{center}
         \includegraphics[width=8.0cm]{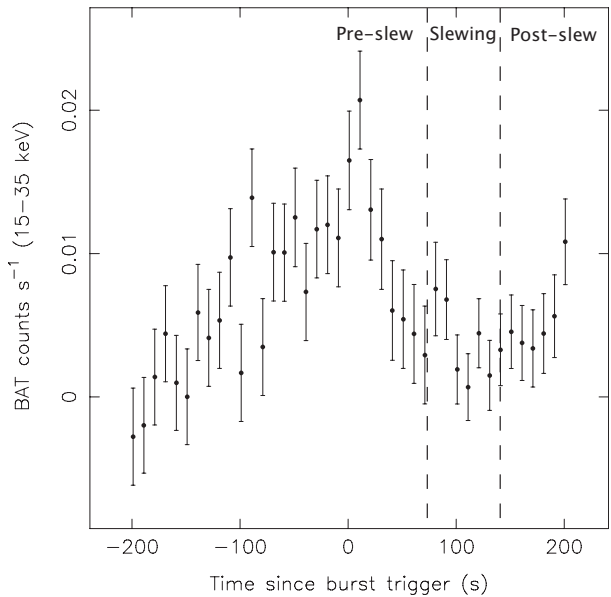}
    \end{center}
\caption[]{{Background subtracted lightcurve of \rxh\ from \swift/BAT data at 10 s time resolution (15--35 keV). The times at which the spacecraft started and finished slewing are indicated by the dashed lines.}}
 \label{fig:bat}
\end{figure}

We generated standard BAT data products for the trigger observation using the \textsc{batgrbproduct} tool. The 15--35 keV BAT lightcurve of the burst, shown in Fig.~\ref{fig:bat}, is consistent with a single peak centred at $t\sim0$~s and emerging from the background for $\sim 200$~s, with a very slow rise time of $\sim 100$~s \citep{baumgartner08,israel08}\footnote{$\mathrm{See~also~http://gcn.gsfc.nasa.gov/notices\_s/311603/BA}$.}. In Fig.~\ref{fig:bat}, the apparent peak at $t\sim90$~s is thought to be an artefact related to the spacecraft slewing, whereas the apparent rise in count rate after $t\sim120$ is likely caused by entering the South Atlantic Anomaly \citep[SAA;][]{baumgartner08}.

The spacecraft started slewing $\sim75$~s after the burst trigger, by which time the BAT count rate had nearly dropped to the background level (see Fig.~\ref{fig:bat}). Therefore, we used only pre-slew data and extracted a single BAT spectrum of 140 s around the burst peak using the tool \textsc{batbinevt}. Given the low count rate, it is not useful to divide the BAT data in multiple bins with a higher time resolution. Necessary geometrical corrections were applied with \textsc{batupdatephakw} and the BAT-recommended systematical error was administered using \textsc{batphasyserr}. We generated a single response matrix by running the task \textsc{batdrmgen} and fitted the BAT spectrum between 15--35 keV with \textsc{xspec} \citep[v. 12.5;][]{xspec}.

About 144 s after the BAT trigger, follow-up observations with the narrow-field XRT and UVOT commenced. These observations typically consists of a number of short data segments ($\lesssim 2$~ks), which represent different satellite orbits.

\subsubsection{XRT}\label{subsubsec:xrt}
\begin{figure}
 \begin{center}
         \includegraphics[width=8.0cm]{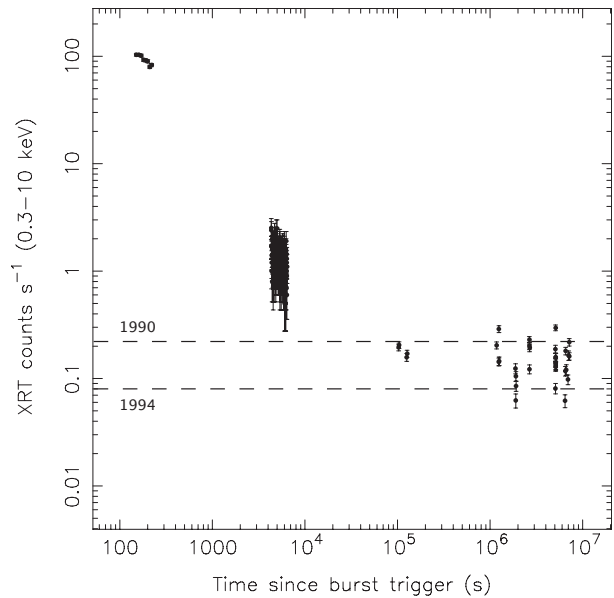}
    \end{center}
\caption[]{{\swift/XRT lightcurve of \rxh\ obtained in 2008. For representation purposes different time bins have been chosen; the first two data sets (respectively WT and PC mode data of observation 311603000) have a time resolution of 10 s, while that of later observations (ID 311603001 onwards) is 1000 s. The two dashed lines represent the intensity levels detected by \rosat\ in 1990 (PSPC) and in 1994 (HRI) converted to XRT count rates (see Section~\ref{subsec:rosat}).}}
 \label{fig:xrt}
\end{figure}

The first XRT data set (ID 311603000) was obtained in windowed timing (WT) mode and consisted of two segments, the first of which lasted for 82 s from 2008 May 14 10:35:05-10:36:27 \textsc{UT}. The source displays a rapid fading during this observation. After a data gap of more than one hour the source was observed for another 8 s from 11:44:40 to 11:44:48 \textsc{UT}. By this time the source count rate had decreased from $\sim 100~\cnts$ down to $\sim 1~\cnts$ (see Fig.~\ref{fig:xrt}), which caused an automatic switch to the photon counting mode (PC). The PC data runs from 2008 May 14 11:44:50 to 12:19:20 \textsc{UT}, amounting to $2068.5$~s of exposure time (ID 311603000). A continued fading is apparent in the X-ray lightcurve of this observation, which suggests that the X-ray burst was ongoing.  

During subsequent observations performed the next day (ID 311603001; WT mode), the source was detected at a count rate of $\sim 0.13~\cnts$. In the following months it remained at that level fluctuating by a factor of $\sim2$ between $\sim 0.06-0.18~\cnts$ (see Table~\ref{tab:persistent}). This indicates that the source had returned to its persistent level the day after the BAT trigger. Fig.~\ref{fig:xrt} displays the lightcurve of all XRT data obtained in 2008. The intensity levels detected with \rosat\ in 1990 and 1994 (see Section~\ref{subsec:rosat}) are also indicated in this plot.\\

To obtain cleaned data products we processed all raw XRT data with the task \textsc{xrtpipeline} using standard quality cuts and selecting event grades 0--12 for the PC mode and 0--2 in the WT mode\footnote{See http://heasarc.gsfc.nasa.gov/docs/swift/analysis for standard \textit{Swift} analysis threads.}.
Source lightcurves and spectra were extracted with \textsc{xselect} (v. 2.3). We used a region of 40 $\times$ 40 pixels to extract source events from the WT data. A region of similar shape and size, positioned on an empty part well outside the point spread function of the source, was used for the background. For the PC mode observations we used a circular region with a 10 pixel radius to extract source photons. An annulus with an inner (outer) radius of 75 (100) pixels, centred on the source position, served as the background reference. We generated exposure maps with the task \textsc{xrtexpomap} and ancillary response files (ARF) were created with \textsc{xrtmkarf}. 
The response matrix files (v. 11; RMF) were obtained from the CALDB database. 

The spectra were grouped using the FTOOL \textsc{grppha} to contain bins with a minimum number of 20 photons. We fitted the spectra with \textsc{xspec} in the 0.5--10 keV range. The PC data of observation 311603000 was affected by pile-up. Following the \swift\ analysis threads\footnote{See http://www.swift.ac.uk/pileup.shtml.\label{foot:pileup}}, we attempted to correct for the consequent effect on spectral shape and loss in source flux by using an annulus with an inner (outer) radius of 4 (10) pixels as the source extraction region.

We performed time-resolved spectroscopy of the fading tail of the X-ray burst using the  XRT observations of May 14 (both WT and PC mode data; ID 311603000). The first set of WT data was divided into 4 intervals of 20 s, each with a total of $\sim 2000$ counts per interval. We do not include the second set of WT data in the analysis, since this 8 s exposure collected only 14 source photons and the consecutive PC data provide better statistics. The $\sim 2$~ks PC mode observation consists of a single data segment, which was split into two intervals of similar length, containing $\sim 500$ counts each after pile-up correction. 

We searched the $\sim90$~s long WT observation of the X-ray burst for periodicities by means of Fast Fourier Transforms (FFTs) and applying the method described in \citet{israel96}. The analysed period range spans from $\sim$3.5~ms up to 100~s ($\sim$262\,000 total period trials) and the Nyquist frequency is $\sim283$~Hz. No significant peaks were found. Meaningful upper limits ($<$100 per cent pulsed fraction) are obtained only for periods shorter than 5~s and range between $\sim$15 and $\sim$20 per cent.

To characterize the persistent emission, we used the data obtained from May 15 onwards (IDs 311603001-- 311603015). The upper left panel of Fig.~\ref{fig:images} displays a summed X-ray image of all PC mode observations of the post-burst epoch. We obtained another \swift/XRT pointing in late July 2009 (ID 31446001) to investigate the state of the system more than a year after the X-ray burst. During that observation, \rxh\ is detected at a count rate of $\sim 0.11~\cnts$. This is the same level as detected in 2008 May--August (see also Table~\ref{tab:persistent}), which indicates that the system is still actively accreting (see Sections 3 and 4).\\

\subsubsection{UVOT}\label{subsubsec:uvot}
\begin{figure*}
 \begin{center}
       \includegraphics[width=8.0cm]{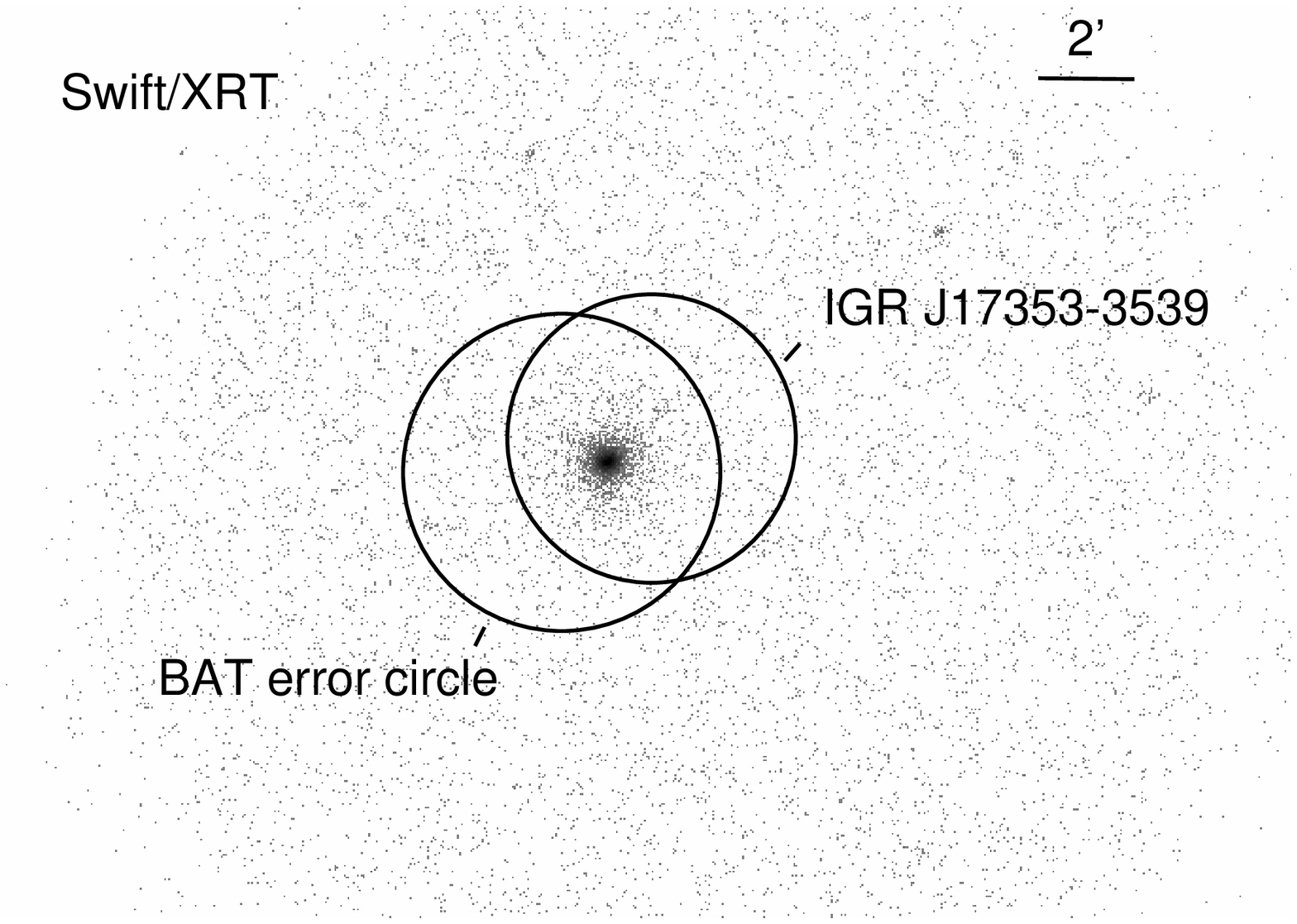}\hspace{0.1cm}
       \includegraphics[width=8.0cm]{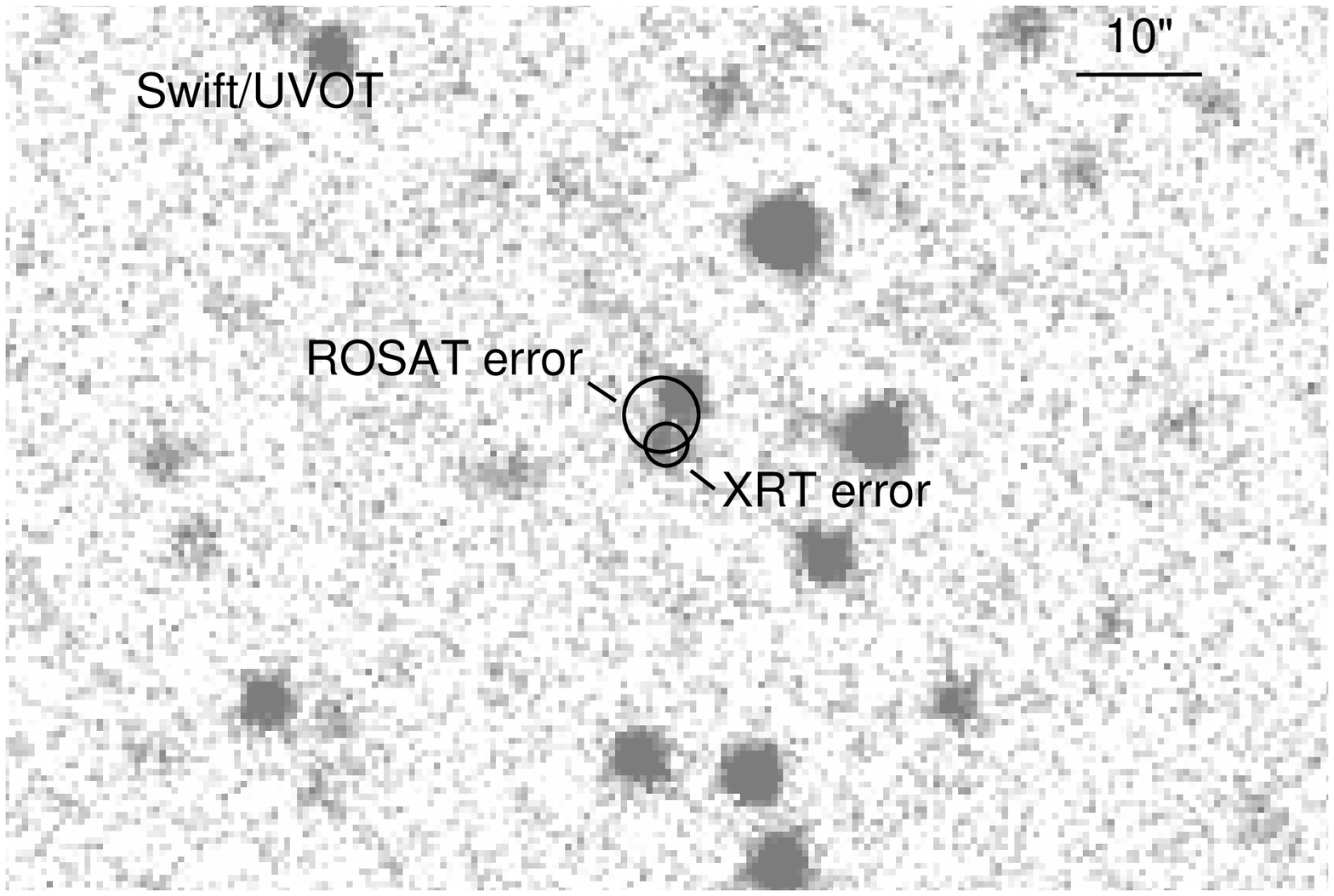}\\ \vspace{0.1cm}
           \includegraphics[width=8.0cm]{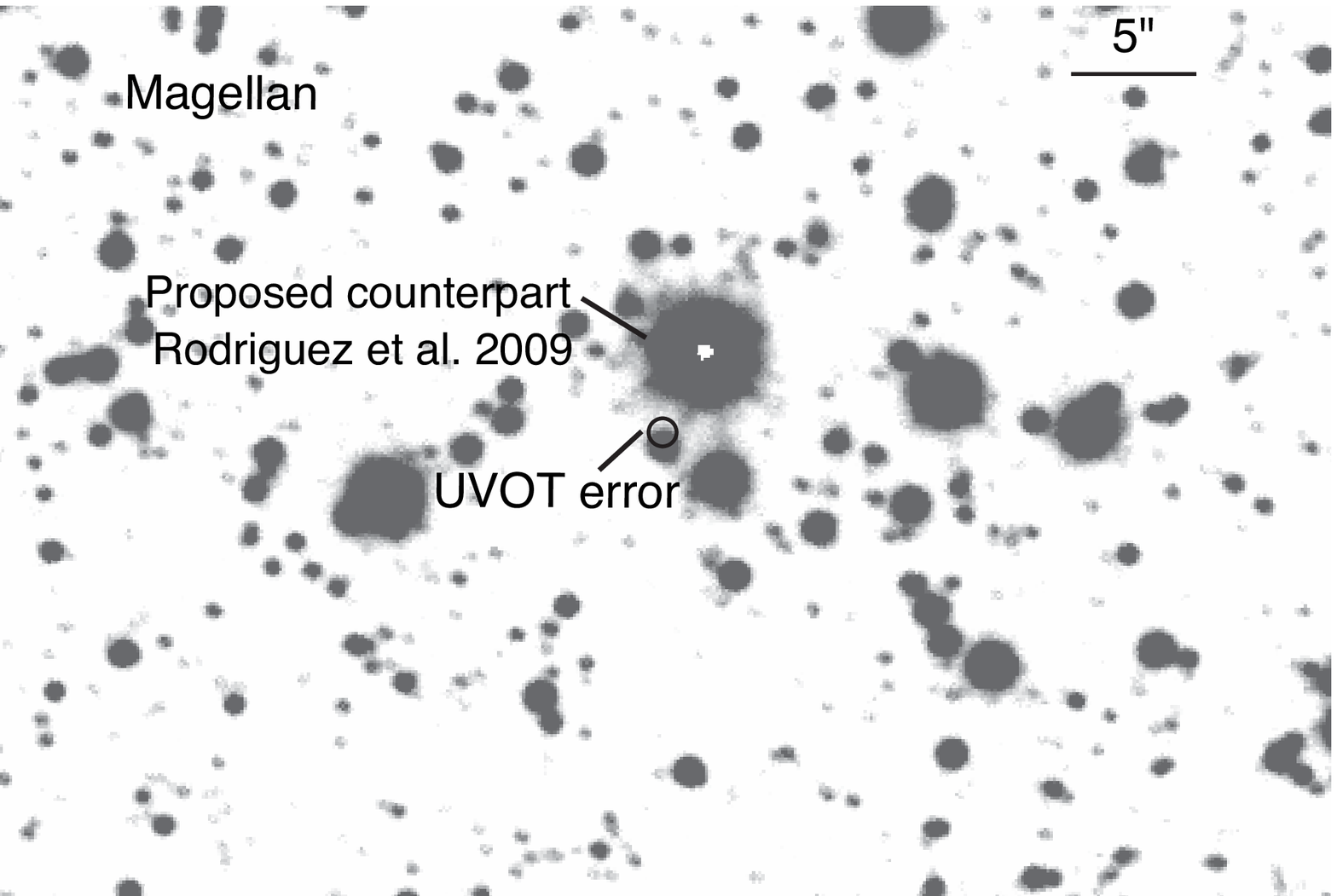}\hspace{0.1cm}
           \includegraphics[width=8.0cm]{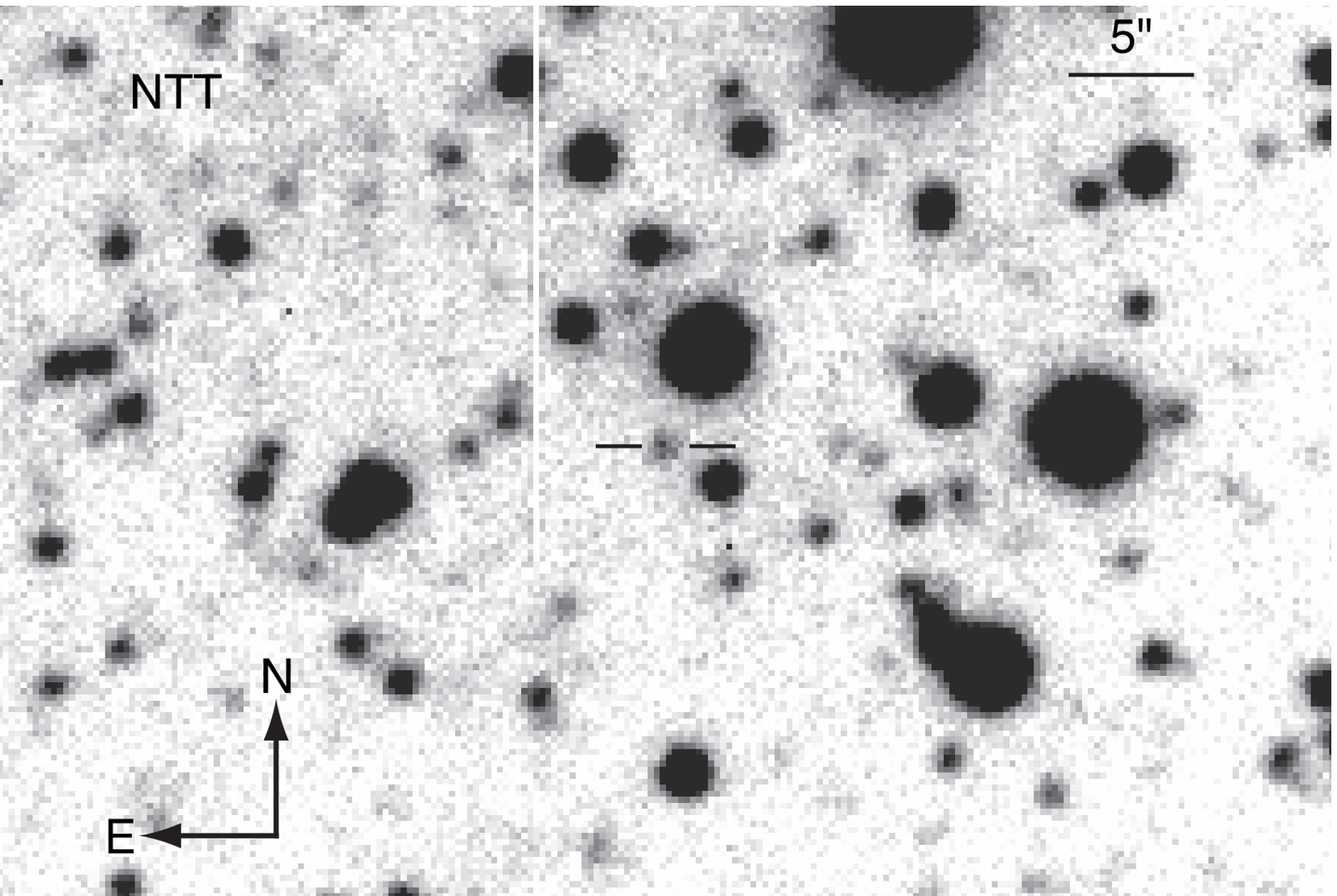}
    \end{center}
\caption[]{{Images of the field around \rxh. Upper left: summed X-ray image of \swift/XRT PC mode data (0.3--10 keV) obtained after the X-ray burst in 2008 (IDs 311603004--311603015). The BAT error circle and the \inte\ position of \igr\ are also indicated (see Section~\ref{subsec:integral}). Upper right: \swift/UVOT $WH$-band image obtained during the X-ray burst (ID 311603000). The 3 arcsec \rosat/HRI error circle (see Section~\ref{subsec:rosat}), as well as the 1.7 arcsec \swift/XRT error circle are indicated in this image. Lower left: \textit{Magellan} $J$-band image. The circle represents a $1\sigma$ error circle around the UVOT position. The counterpart proposed by \citet{rodriguez09} is also indicated (see Section~\ref{sec:intro}). Lower right: $V$-band optical image obtained with the \ntt.}}
 \label{fig:images}
\end{figure*}

The UVOT data of \rxh\ were obtained using a variety of filters, but the source could only be detected in the broadband white filter (\textit{WH}, $\sim 1500-8500$ \AA{}). The upper right panel of Fig.~\ref{fig:images} shows an UVOT $WH$-band image of the field around \rxh\ and Table~\ref{tab:obs} gives an overview of the UVOT observations obtained with this filter.

Avoiding a nearby object (see Fig.~\ref{fig:images}), we used a circular region with a radius of 2 arcsec to extract source photons, and a source-free region with a radius of 10 arcsec as a background reference. Magnitudes were extracted using the tool \textsc{uvotsource}, taking into account aperture corrections.

During the X-ray burst decay, there were three intervals of UVOT observations using the $WH$-filter. The bottom panel of Fig.~\ref{fig:burst_fit} shows the evolution of the magnitude during these intervals; there is a clear decay visible (two magnitudes within two hours), simultaneous with the observed fading in X-rays. This provides strong evidence that the fading UVOT source represents the optical counterpart of the system and allows for a sub-arcsecond localization of the burster \citep{israel08}.\\

\subsection{Ground-based optical/near-IR photometry}
All optical photometric observations discussed in this section were reduced using standard routines in \textsc{iraf}\footnote{\textsc{iraf} is distributed by the National Optical Astronomy Observatories, which are operated by the Association of Universities for Research in Astronomy, Inc., under cooperative agreement with the National Science Foundation.} by subtracting an average bias frame and dividing by a normalized flat field. The near-IR data was reduced using \textsc{iraf} and the specific PANIC package provided by the Las Campanas Observatory.

\subsubsection{REM}\label{subsubsec:rem}
The \rem\ is a 60-cm fast slewing telescope located at la Silla, Chile, which is dedicated to prompt optical/near-IR follow-ups of GRB afterglows \citep[][]{zerbi01,chincarini03,covino04}. The \rem\ automatically responded when BAT triggered on the X-ray burst from \rxh\ and started observing with the ROSS camera 188 s after the BAT trigger. A series of 5 \textit{R}-band observations with exposure times of 30 s were carried out during the X-ray burst decay and two more frames of 120 s each were obtained the next day, all in the \textit{R}-band. These images show a source declining in brightness at a position consistent with the UVOT location. The seeing during the observations was 1\farcs7 and 1\farcs8 on 2008 May 14 and 15 respectively.

Astrometry was performed using the USNOB1.0\footnote{http://www.nofs.navy.mil/data/fchpix.} catalogue and aperture photometry was done with the \textsc{SExtractor} package \citep{bertin96} for all the objects in the field. The calibration was done against Landolt standard stars. In order to minimize any systematic effect, we performed differential photometry with respect to a selection of local isolated and non-saturated standard stars.

\subsubsection{Magellan}\label{subsubsec:magellan}
\citet{cackett08} already reported on near-IR observations of the field around \rxh, carried out with the PANIC camera \citep{panic04} on the 6.5-m \textit{Magellan Baade} telescope. We summarize those observations here. On 2008 May 25, eleven days after the occurrence of the X-ray burst, images were acquired in the $J$-, $H$- and $Ks$-bands for total on-source times of 600, 300 and 300 s respectively. The observations were taken in a series of 5 pattern dithers; the separate images were shifted and combined in the standard way. The astrometry was tied to known 2MASS sources in the field, which were also used to calibrate the photometry of \rxh.  

In all three bands a source is detected at a position of $\alpha= \mathrm{17^{h}35^{m}23.74^{s}}$, $\delta=-35^{\circ} 40' 16.6\arcsec$ (J2000) with an uncertainty of 0.1 arcsec. Within the errors, this is consistent with the UVOT coordinates of \rxh, implying that this is the possible near-IR counterpart. The lower left panel of Fig.~\ref{fig:images} shows the \textit{J}-band image.

\subsubsection{NTT}\label{subsubsec:ntt}
Further optical photometric observations were performed on 2008 June 16, using the EFOSC2 camera on the ESO 3.6-m \ntt\ located at la Silla. Images were obtained in the $B$-, $V$- and $R$-waveband for total exposure times of 900, 900 and 2100 s respectively. During these observations the seeing was varying between 0\farcs9 and 1\farcs4. 

In both the $V$- and the $R$-band, a weak source is detected right at the position of the near-IR source found in \magellan\ images. The field around \rxh\ was calibrated against SA110 Landolt standard field stars that were observed on the same night. We corrected the instrumental magnitudes using the average atmospheric extinction mentioned on the La Silla website\footnote{http://www.eso.org/sci/facilities/lasilla/telescopes/d1p5/misc/\\Extinction.html.}.

\begin{table*}
\begin{center}
\caption[l]{Results from spectral analysis of the post-burst \swift/XRT data (IDs 311603001--31446001).}
\begin{threeparttable}
\begin{tabular}{l l l l l l l l l}
\hline \hline
Date & Mode & Count rate & $\Gamma$ & $F_{\mathrm{X}}^{\mathrm{abs}}$ & $F_{\mathrm{X}}^{\mathrm{unabs}}$ & $L_{\mathrm{X}}$ \\
- & - & ($\cnts$) & -& ($10^{-11}~\mathrm{erg~cm}^{-2}~\mathrm{s}^{-1}$) & ($10^{-11}~\mathrm{erg~cm}^{-2}~\mathrm{s}^{-1}$) & ($10^{35}~\mathrm{erg~s}^{-1}$) \\
\hline
2008-05-15 & WT & 0.13 & $2.2\pm0.3$ & $0.77\pm0.09$ & $1.5\pm0.2$ & $1.6\pm0.2$  \\ 
2008-05-15 & WT & 0.08 & $2.4\pm0.4$ &  $0.51\pm0.07$ &  $1.2\pm0.3$ & $1.3\pm0.3$  \\ 
2008-05-28 & PC & 0.17 & $2.1\pm0.2$ & $1.20\pm0.10$ & $2.2\pm0.3$ & $2.4\pm0.3$ &  \\
2008-06-05 & PC & 0.08 & $2.5\pm0.3$ & $0.53\pm0.04$ &  $1.2\pm0.2$ & $1.3\pm0.2$  \\
2008-06-14 & PC & 0.18 & $2.2\pm0.2$ & $1.30\pm0.10$ & $2.5\pm0.3$ & $2.7\pm0.3$  \\
2008-07-12 & PC & 0.14 & $2.1\pm0.2$ &$0.92\pm0.03$ & $1.7\pm0.2$ &  $1.8\pm0.2$ \\
2008-07-28 & PC & 0.06  &$3.0\pm0.6$ & $0.33\pm0.05$ & $1.2\pm0.5$ & $0.7\pm0.3$\\
2008-07-29 & PC & 0.18 &  $2.3\pm0.2$ & $1.10\pm0.10$ & $2.2\pm0.3$ & $2.4\pm0.3$  \\
2008-07-31 & PC & 0.07  & $2.4\pm0.8$ & $0.45\pm0.09$ & $1.0\pm0.4$ & $1.1\pm0.4$ \\
2008-08-02 & PC & 0.08  & $2.5\pm0.6$ & $0.42\pm0.10$ & $1.0\pm0.3$ & $1.1\pm0.3$ \\
2008-08-05 & PC & 0.17  & $2.5\pm0.3$ & $0.99\pm0.08$ & $2.3\pm0.4$ &  $2.5\pm0.4$ \\
2009-07-24 & PC & 0.11 & $2.5 \pm 0.3$ & $0.73 \pm 0.08$ & $1.7 \pm 0.2$ & $1.8\pm0.2$ \\
\hline
\end{tabular}
\label{tab:persistent}
\begin{tablenotes}
\item[] \textit{Note}. The quoted errors represent 90 per cent confidence levels. The hydrogen column was tied between the observations; the best fit yielded $N_{\mathrm{H}}=(9.3\pm1.0)\times10^{21}~\mathrm{cm}^{-2}$ for a reduced $\chi^2=1.1$ (190 d.o.f.). The quoted fluxes are in the 0.5--10 keV energy range and the luminosity in that band was calculated assuming a distance of $D=9.5$~kpc.
\end{tablenotes}
\end{threeparttable}
\end{center}
\end{table*}

\subsection{Optical spectroscopy}\label{subsec:vlt}
Through a DDT request we obtained three 1200~s long slit spectra on 2008 July 27 00:29 -- 01:10 UT with the FORS2 instrument mounted on the 8.2-m \vlt. We used the 600RI holographic grism, a slit width of 1.0 arcsec and the CCD detector binned by 2 to provide a dispersion of 1.63~\AA\ per pixel in the wavelength range $\lambda\lambda 5300-8600$. The observations took place under a 0\farcs7 seeing, yielding a spectral resolution of $220$ and $160~\kms$ at \halfa\ and 8500~\AA\ respectively.

The spectra were reduced using the \textsc{iraf kpnoslit} package. The data were bias subtracted, flat field corrected and optimally extracted \citep{horne1986}. Wavelength calibration was performed using lines from He, HgCd, Ar \& Ne lamp spectra obtained with the same instrumental set-up during daytime, the day after the observations -- as is customary for \vlt\ service mode observations. The extracted spectra were analysed further using the \textsc{iraf} tool {\sc splot} and the software package \textsc{molly}.

\subsection{Flux history}\label{subsec:rosat}

\subsubsection{\rosat}
\rxh\ was discovered in 1990 September during an all sky-survey with the Position Sensitive Proportional Counter (PSPC) aboard the \rosat\ satellite (ID RS932341). Pointed follow-up observations with the High Resolution Imager (HRI) were carried out on 1994 October 1 (ID RH900607). According to the \rosat\ online catalogue\footnote{http://www.xray.mpe.mpg.de/cgi-bin/rosat/src-browser.} the detected count rates were $0.14\pm 0.02~\cnts$ for the PSPC and $0.021\pm 0.004~\cnts$ for the HRI (0.1--2.5 keV). 

Employing \textsc{pimms} and adapting the spectral parameters found for the persistent X-ray emission of \rxh\ (see Section~\ref{subsec:persistent}; $N_{\mathrm{H}}=9.3 \times 10^{21}~\mathrm{cm}^{-2}$ and a powerlaw index $\Gamma=2.3$), this translates into 0.5--10 keV unabsorbed fluxes of $(1.8\pm0.3) \times 10^{-11}~\flux$ and $(7.2 \pm 1.4)\times 10^{-12}~\flux$, for the PSPC and HRI respectively. The corresponding \swift/XRT PC mode count rates are $0.18 \pm 0.03$ and $0.07 \pm 0.01~\cnts$, consistent with the persistent emission detected with \swift/XRT in 2008 and 2009 (see Fig.~\ref{fig:xrt} and Table~\ref{tab:persistent}).

\subsubsection{\xmm}
In addition to the above mentioned \rosat\ detections, \rxh\ was observed with \xmm\ on 2008 March 4, which is 10 weeks prior to the X-ray burst caught by \swift, as part of the \xmm\ slew survey \citep[][]{xmmslew05}\footnote{\rxh\ appears in the third update of the catalogue, which was released in July 2009, and is assigned the name XMMSL1 J173524.0--354021.}. The source was detected with the European Photon Imaging Camera (EPIC) PN instrument at a count rate of $2.24 \pm 0.53~\cnts$ (0.2--12 keV), which converts into a 0.5--10 keV unabsorbed flux of $\sim(4 \pm 1) \times 10^{-11}~\flux$ (again using \textsc{pimms} with $N_{\mathrm{H}}=9.3 \times 10^{21}~\mathrm{cm}^{-2}$ and a powerlaw index $\Gamma=2.3$). The corresponding \swift/XRT count rate is $\sim 0.40 \pm 0.10~\cnts$.

\subsection{Searches for other X-ray bursts}\label{subsec:integral}

\subsubsection{\inte}
\rxh\ lies within the 3 arcmin error box of the unclassified hard X-ray source \igr\ \citep[see Fig.~\ref{fig:images}; this coincidence was also noted by][]{rodriguez09}, which appears in the \inte\ all-sky survey catalogue \citep{krivonos07,bird09}.

We used the publicly available \inte\ data to search for X-ray bursts from the location of \rxh/\igr. This region has been covered by regular observations of the \inte\ satellite \citep{winkler2003} since the beginning of 2003, in particular at low energy (3--20 keV) with the Joint European X-ray Monitor \citep[JEM-X;][]{lund2003}, module 1 and 2, and at high energies (17--100 keV) with the \inte\ Soft Gamma-ray Imager \citep[ISGRI;][]{lebrun2003}, mounted on the Imager onBoard the \inte\ Satellite \citep[IBIS;][]{ubertini2003}. The data are divided into individual pointings called Science Windows (ScW), themselves grouped into revolutions of the satellite. \inte\ was not pointing towards the source field when the X-ray burst picked up by BAT occurred. 

In the archival public data, there are 7359 IBIS ScW between revolutions 37 and 674, pointing less than 12 degrees from the source, and 650 JEM-X ScW between revolutions 46 and 661, pointing less than 3.5 degrees from the source. These data are spread over a time range of five years, from 2003 February 1 to 2008 April 20, for effective exposures of 16 and 0.76~Ms for IBIS and JEM-X respectively. The difference of exposure is due to the fact that IBIS has a larger field of view than JEM-X and thus happened to observe \igr\ more often.

We have analysed this data set with the standard Offline Science Analysis software (OSA; v. 7.0), distributed by the \inte\ Science Data Center \citep[ISDC;][]{courvoisier2003} and based on algorithms described in \citet{goldwurm2003} for IBIS and \citet{westergaard2003} for JEM-X.
The total collapsed mosaic of the IBIS images reveals a weak but significant (7.7$\sigma$) excess at the position of the source. Its flux in the 17--40 keV band is $\sim 4 \times 10^{-12}~\flux$. 
We have searched for X-ray bursts in the IBIS data with the \inte\ Burst Alert System \citep[IBAS;][]{mereghetti2003}, yet no X-ray burst was detected. We have also explored the JEM-X data, more suitable to look for such events since these are usually soft. However, again, no X-ray burst was found.

\subsubsection{\swift/BAT}
We investigated the \swift/BAT transient monitor results of \rxh, provided by the \swift/BAT team\footnote{See http://swift.gsfc.nasa.gov/docs/swift/results/transients/\\weak/1RXHJ173523.7--354013.}. No other X-ray bursts are detected with a limiting flux of $\sim1.4 \times 10^{-9}~\flux$ (15--50 keV) for a single pointing (which have a mean duration $\sim700$~s). However, the energy range of the BAT transient monitor (15--50 keV) is not optimally sensitive to X-ray bursts as soft as the May 14 event and it is therefore possible that an X-ray burst brighter than $\sim1.4 \times 10^{-9}~\flux$ has been missed in this wider band. During the five years of the \swift\ mission there have been no other onboard triggers comparable in intensity to the X-ray burst of 2008 May 14. The total BAT exposure time till 2009 August 5 is 4.3 Ms.

 \begin{figure}
 \begin{center}
    \includegraphics[width=8.0cm]{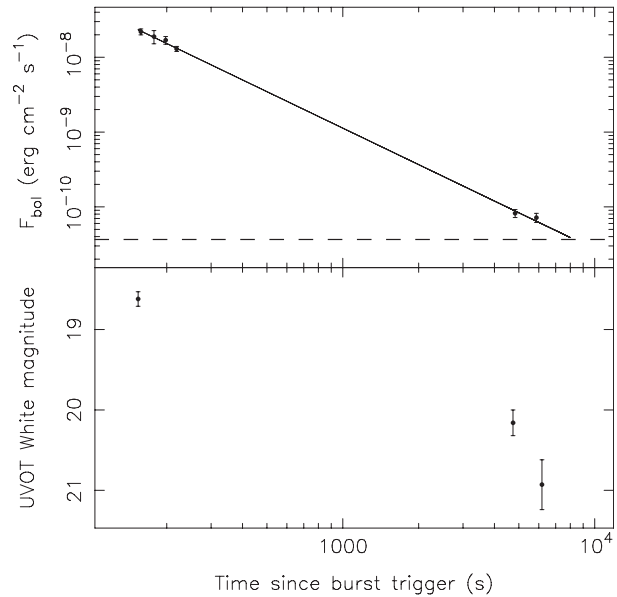}
    \end{center}
\caption[]{{Top: bolometric flux of the cooling tail of the X-ray burst (XRT data) along with a powerlaw fit with index -1.53. The dashed line in the top panel indicates the estimated persistent bolometric flux. Bottom: evolution of the UVOT $WH$-filter magnitude during the X-ray burst.}}
 \label{fig:burst_fit}
\end{figure}

\section{Results}

\subsection{Persistent X-ray emission}\label{subsec:persistent}
The post-burst data taken with \swift/XRT (see Table~\ref{tab:persistent} for an overview) was modelled with an absorbed powerlaw continuum modified by absorption (\textsc{phabs}; we used the default \textsc{xspec} abundances and cross section for this model). We fitted all spectra simultaneously with the hydrogen column density tied between all observations. The results of this simultaneous modelling, which yielded a final reduced $\chi^2=1.1$ for 190 degrees of freedom (d.o.f.), are presented in Table~\ref{tab:persistent}. The values of the spectral parameters were not significantly different when each observation was fit separately. 

The best fit hydrogen column density is $N_{\mathrm{H}}=(9.3 \pm 1.0)\times10^{21}~\mathrm{cm}^{-2}$ and the powerlaw index is consistent with being constant within the spectral errors (the average value is $\Gamma=2.3 \pm 0.2$). In the days--months following the X-ray burst the source settled at an average unabsorbed flux of $F_{\mathrm{X}}^{\mathrm{unabs}}\sim 1.9 \times 10^{-11}~\flux$ (0.5--10 keV), varying by a factor of $\sim 2$. Assuming a bolometric correction factor of 2 \citep[][]{zand07}, we estimate a bolometric persistent flux of $F_{\mathrm{bol}}^{\mathrm{pers}}\sim 3.8 \times 10^{-11}~\flux$ (0.01--100 keV), which equals $\sim0.1$~per~cent of Eddington for a distance of 9.5 kpc.

\begin{table*}
\begin{center}
\caption[l]{Results from spectral analysis of the X-ray burst data (ID 311603000).}
\begin{threeparttable}
\begin{tabular}{l l l l l l}
\hline \hline
$\Delta T$ & Mode & $F_{\mathrm{bol}}$  & $kT_{\mathrm{bb}}$ &  $R_{\mathrm{bb}}$ & Reduced $\chi^{2}$ (d.o.f.) \\
(s) & - & ($10^{-8}~\flux$) &  (keV) &  (km) & - \\
\hline
70 & BAT & $3.5^{+5.0}_{-1.7}$ & $2.3^{+0.5}_{-0.4}$ & $9.9^{+1.2}_{-0.5}$ & 1.3 (9)   \\ 
10 & WT & $2.2\pm0.2$ & $2.2\pm0.1$ & $9.2\pm0.3$ & 1.0 (91)  \\ 
10 & WT & $1.9\pm0.2$ & $2.1\pm0.1$ & $9.3\pm0.3$ & 1.4 (86) \\ 
10 & WT & $1.8\pm0.2$ & $2.0\pm0.1$ & $9.2\pm0.3$ & 1.2 (79)  \\ 
11 & WT & $1.4\pm0.1$ & $1.9\pm0.1$ &  $9.3\pm0.3$ & 1.6 (77) \\ 
500 & PC & $(1.1\pm0.1)\times10^{-2}$ & $0.67\pm0.04$ & $6.8\pm0.5$ & 1.5 (24) \\ 
535 & PC & $(1.0\pm0.1)\times10^{-2}$ & $0.64\pm0.04$ & $6.9\pm0.6$ & 0.6 (21) \\ 
\hline
\end{tabular}
\label{tab:burst}
\begin{tablenotes}
\item[] \textit{Note}. The quoted errors represent 90 per cent confidence levels. The hydrogen column density was fixed at $N_{\mathrm{H}}=9.3\times10^{21}~\mathrm{cm}^{-2}$ and a distance of $D=9.5$~kpc was used to calculate the emitting radius from the model normalization. $\Delta T$ indicates the size of the time interval.
\end{tablenotes}
\end{threeparttable}
\end{center}
\end{table*}

\subsection{X-ray burst}\label{subsec:burst_ana}

\subsubsection{Spectra and ligthcurve}
We fitted the BAT (15--35 keV) and XRT (0.5--10 keV) X-ray burst spectra with an absorbed blackbody model \textsc{bbodyrad}, which has a normalization that equals $R_{\mathrm{bb}}^2/D_{10}^2$, where $R_{\mathrm{bb}}$ is the emitting radius in km and $D_{10}$ is the source distance in units of 10 kpc. We kept the hydrogen column density fixed at the value found from fitting the persistent emission spectra ($N_{\mathrm{H}}=9.3 \times10^{21}~\mathrm{cm}^{-2}$). Since the last two XRT data segments trace the faint end of the X-ray burst, the underlying persistent emission must be taken into account. Therefore, we add a powerlaw component in the spectral fits, for which the index and normalization are fixed at the average values found from modelling the persistent emission (see Section~\ref{subsec:persistent}). To estimate the bolometric fluxes during the X-ray burst, we extrapolate the fitted blackbody component to the 0.01--100 keV energy range. For the BAT data we find a blackbody temperature of $kT_{\mathrm{bb}}=2.3^{+0.5}_{-0.4}$~keV and an unabsorbed 0.01--100 keV flux of $F_{\mathrm{bol}}^{\mathrm{peak}}=3.5^{+5.0}_{-1.7} \times 10^{-8}~\flux$. 

X-ray bursts picked up by BAT are typically the most energetic bursts, which frequently show photospheric radius-expansion (PRE) indicating that the Eddington luminosity is reached during the burst peak. However, the low number of counts in the BAT data and the gap with the XRT observations preclude a spectral confirmation of such an expansion (i.e., a local peak in emitting radius associated with a dip in blackbody temperature).
If we assume that the peak flux was equal to or lower than that typical of PRE bursts \citep[$3.8 \times 10^{38}~\lum$;][]{kuulkers03_xrb}, we can place an upper limit on the source distance of $D\lesssim9.5$~kpc. However, for a H-rich photosphere (H-fraction X=0.7), the empirically derived Eddington limit is $1.6 \times 10^{38}~\lum$ \citep[][]{kuulkers03_xrb} and this would lower the distance estimate to $D\lesssim6.2$~kpc. In this work we have adopted a distance of 9.5 kpc when calculating luminosities, energies and blackbody emitting radii.

\begin{figure}
 \begin{center}
    \includegraphics[width=8.0cm]{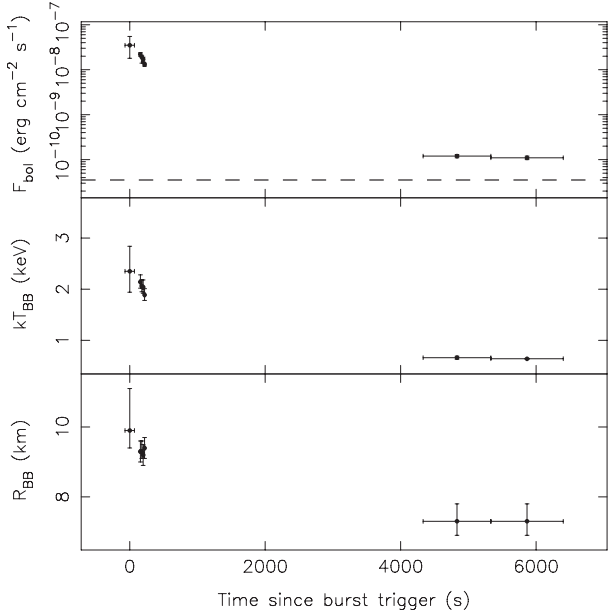}
    \end{center}
\caption[]{{Results from time-resolved spectroscopy of the \swift\ BAT (first point) and XRT burst data. The dashed line in the top panel corresponds to the estimated persistent bolometric flux.}}
 \label{fig:burst_spec}
\end{figure}

The results from our time-resolved spectroscopic analysis of the BAT and XRT data are presented in Table~\ref{tab:burst} and Fig.~\ref{fig:burst_spec}. The 0.01--100 keV flux decreases by 2 orders of magnitude from $F_{\mathrm{bol}}^{\mathrm{peak}}\sim3.5 \times 10^{-8}~\flux$ at the time of the BAT trigger down to $F_{\mathrm{bol}}^{\mathrm{}}\sim 1.0 \times 10^{-10}~\flux$ in the final XRT data interval, which was obtained 1.6 h after the burst trigger. The blackbody temperature decreases from $kT_{\mathrm{bb}}\sim2.3$~keV at the peak down to $kT_{\mathrm{bb}}\sim0.6$~keV in the tail of the X-ray burst.

Based on theoretical modelling, the flux in the cooling tails of long X-ray bursts is expected to follow a powerlaw decay \citep{cumming04}. The XRT light curve data, representing the burst tail, can be fit with a simple powerlaw with index $-1.53 \pm 0.03$ and a normalization of $(5.3\pm1.0)\times 10^{-5}~\flux$ (reduced $\chi^2=1.0$ for 2 d.o.f.). Extrapolating this fit down to the level of the persistent emission ($F_{\mathrm{bol}}^{\mathrm{pers}}\sim3.8 \times 10^{-11}~\flux$), we can estimate a burst duration of $\sim8000$~s ($\sim2.2$~h; see Fig.~\ref{fig:burst_fit}). A single exponential decay does not provide an adequate fit to the tail of the X-ray burst (reduced $\chi^2\sim8$ for 3 d.o.f.).

\subsubsection{Energetics and ignition conditions}
The X-ray burst was visible in the BAT lightcurve for $\sim 200$~s and we estimated a bolometric flux of $\sim 3.5 \times 10^{-8}~\flux$ from the BAT spectrum (see Table~\ref{tab:burst}). This implies a fluence of $f_{\mathrm{BAT}}\sim 7 \times 10^{-6}~\fluence$ for the BAT data. To estimate the fluence in the burst tail, we integrate the XRT data along the above described powerlaw decay from $t=100$~s (the time at which the burst peak had disappeared from the BAT lightcurve) till $t=8000$ s after the burst trigger (when the flux had decayed down to the persistent level). This way we find a bolometric fluence of $f_{\mathrm{XRT}}\sim 7.7 \times 10^{-6}~\fluence$. The total estimated bolometric fluence of the X-ray burst thus adds up to $f_{\mathrm{burst}}\sim 1.5 \times 10^{-5}~\fluence$. Using the distance upper limit of 9.5 kpc, this implies a maximum radiated energy of $E_{\mathrm{burst}} \lesssim 1.6 \times 10^{41}$~erg. This is more energetic than typical type-I X-ray bursts (see Section~\ref{sec:discuss}).

Using the observed burst energetics, we can calculate the depth at which the X-ray burst ignited. The ignition column depth is given by $y=E_{\mathrm{burst}}(1+z)/4 \pi R^2 Q_{\mathrm{nuc}}$, where $z$ is gravitational redshift, $R$ is the neutron star radius and $Q_{\mathrm{nuc}}=1.6+4$X MeV nucleon$^{-1}$ the nuclear energy release given a H-fraction X at ignition \citep[e.g.,][]{galloway06}. Assuming a neutron star with $M=1.4~\Msun$ and $R=10$~km (so that $z=0.31$), we find an ignition depth of $y\sim 1.5 \times 10^{10}~\mathrm{g~cm}^{-2}$ for pure He (X=0) or $y\sim 5.4 \times 10^{9}~\mathrm{g~cm}^{-2}$ for solar abundances (X=0.7). 

Next we can estimate the recurrence time that corresponds to these ignition depths. A distance of $D\lesssim9.5$~kpc would convert the bolometric persistent flux (see Section~\ref{subsec:persistent}) into a luminosity of $L_{\mathrm{bol}}^{\mathrm{pers}}\lesssim 4.1 \times 10^{35}~\lum$. For a neutron star of mass $M=1.4~\Msun$ and radius $R=10$~km this implies a global mass-accretion rate of $\dot{M} \sim R L_{\mathrm{bol}}^{\mathrm{pers}}/GM \lesssim3.6 \times 10^{-11}~\Msun~\mathrm{yr}^{-1}$ ($\sim0.1$ per cent of the Eddington rate). Assuming isotropy, this corresponds to a local accretion rate (i.e., per unit area) of $\dot{m}\lesssim 1.7 \times 10^2~\mathrm{g~cm}^{-2}~\mathrm{s}^{-1}$. 
Given this local accretion rate and the ignition conditions calculated above, we can estimate the time required to build up the layer that caused the X-ray burst observed from \rxh. We find $t_{\mathrm{rec}} \sim y(1+z)/\dot{m} \gtrsim$ 3.7 yr (X=0) or 1.3 yr (X=0.7). Such a long recurrence time is consistent with the fact that no other X-ray bursts were detected in the entire sample of \inte\ observations (JEM-X and IBIS/ISGR; 16.8 Ms) and the \swift/BAT transient monitor (4.3 Ms).

\subsection{Optical/near-IR photometry}\label{subsec:photometry}
Table~\ref{tab:photometry} summarizes the results from optical and near-IR photometry carried out with different instruments. 
During the X-ray burst, three UVOT $WH$-filter images were obtained, which show a clear fading from a magnitude of $18.6\pm0.1$, 154 s after the burst trigger, down to $20.9\pm0.3$ mag more than an hour later (see Fig.~\ref{fig:burst_fit}). 
The \rem\ telescope acquired two series of $R$-band images. The first (starting 188 s after the BAT trigger) was carried out during the decay of the X-ray burst and clearly detected the source at $R=15.9 \pm 0.2$. During the second set of observations obtained one day later, no source could be detected with a limiting magnitude of $R>17.5$, indicating that the $R$-band flux had faded by $>1.5$~mag. This result is consistent with the decrease in flux observed in the X-ray band.

Within a month after the BAT detection of the X-ray burst from \rxh, optical and near-IR observations were obtained to characterize the persistent emission of the system. In the $J$-, $H$- and $Ks$-band images obtained with \magellan, a source consistent with the UVOT position of \rxh\ was detected (see Fig.~\ref{fig:images}). The \ntt\ observations detected a possible optical counterpart in both the $V$- (see Fig.~\ref{fig:images}) and $R$-band, but no source was detected in the $B$-band. The $R$-band magnitude derived from the \ntt\ observations is consistent with the upper limit obtained from the \rem\ images. The observed apparent magnitudes and colours are listed in Table~\ref{tab:photometry}.

Using the hydrogen absorption column found from fitting the spectral X-ray data ($N_{\mathrm{H}}=9.3 \times10^{21}~\mathrm{cm}^{-2}$), we can calculate the visual extinction. We use the standard relation $N_{\mathrm{H}}/A_{\mathrm{V}} = (1.79 \pm 0.03)\times 10^{21}~\mathrm{atoms~cm}^{-2}$~mag$^{-1}$, which yields $A_{\mathrm{V}}=5.2\pm0.6$~mag \citep{predehl1995}. The extinction in the other bands can be estimated using the relations $A_{\mathrm{B}}/A_{\mathrm{V}}$=1.325, $A_{\mathrm{R}}/A_{\mathrm{V}}$=0.748, $A_{\mathrm{J}}/A_{\mathrm{V}}$=0.282, $A_{\mathrm{H}}/A_{\mathrm{V}}$=0.175 and $A_{\mathrm{K}}/A_{\mathrm{V}}$=0.112 \citep{rieke1985}. The de-reddened magnitudes and colours are also listed in Table~\ref{tab:photometry}.

\begin{table}
\begin{center}
\caption[l]{Apparent magnitudes and colours derived from optical/near-IR photometry.}
\begin{threeparttable}
\begin{tabular}{l l l l}
\hline \hline
Date & Band  & Observed & De-reddened \\
\hline
\multicolumn{4}{l}{\textbf{X-ray burst}} \\
2008-05-14 & $WH$ & $18.6 \pm 0.1$  & \\
2008-05-14 & $WH$ & $20.2\pm0.2$ &  \\
2008-05-14 & $WH$ & $20.9\pm0.3$  & \\
2008-05-14 & $R$ & $15.9 \pm0.2$ & $11.7 \pm 0.5$\\
& & & \\
\hline
\multicolumn{4}{l}{\textbf{Persistent emission}} \\
2008-05-15 & $R$ & $>17.5$ & $>13.6$\\
2008-06-16 & $B$ & $>23$ & $>16.1$ \\
2008-06-16 & $V$ & $21.2\pm0.1$ & $16.0 \pm0.6$ \\
2008-06-16 & $R$ & $18.8\pm0.1$ & $14.9 \pm0.5$ \\
2008-05-25 &  $J$ & $15.4 \pm0.1$ & $13.9 \pm0.2$ \\
2008-05-25 &  $H$ & $14.3\pm0.1$ & $13.4 \pm0.1$ \\ 
2008-05-25 &  $K$ & $13.8\pm0.1$ & $13.2 \pm0.1$ \\
&  &  &  \\ 
\hline
\multicolumn{4}{l}{\textbf{Colours}} \\   
2008-06-16 & &  & $0.0\lesssim (V-R)_0 \lesssim 2.2$ \\
2008-06-16 &  &  & $(B-V)_0\gtrsim-0.5$ \\
2008-05-25 &  & &  $0.2\lesssim (J-H)_0 \lesssim 0.8$ \\
2008-05-25 &  &  &  $0.0\lesssim (H-K)_0 \lesssim 0.4$ \\
\hline
\end{tabular}
\label{tab:photometry}
\begin{tablenotes}
\item[] \textit{Note}. The quoted errors and upper limits for the magnitudes represent $1\sigma$ and $3\sigma$ confidence levels respectively. The de-reddened colours represent a $1 \sigma$ range.
\end{tablenotes}
\end{threeparttable}
\end{center}
\end{table}

\subsection{Optical spectra}\label{subsec:specres}
\begin{figure}
 \begin{center}
     \includegraphics[width=8.0cm]{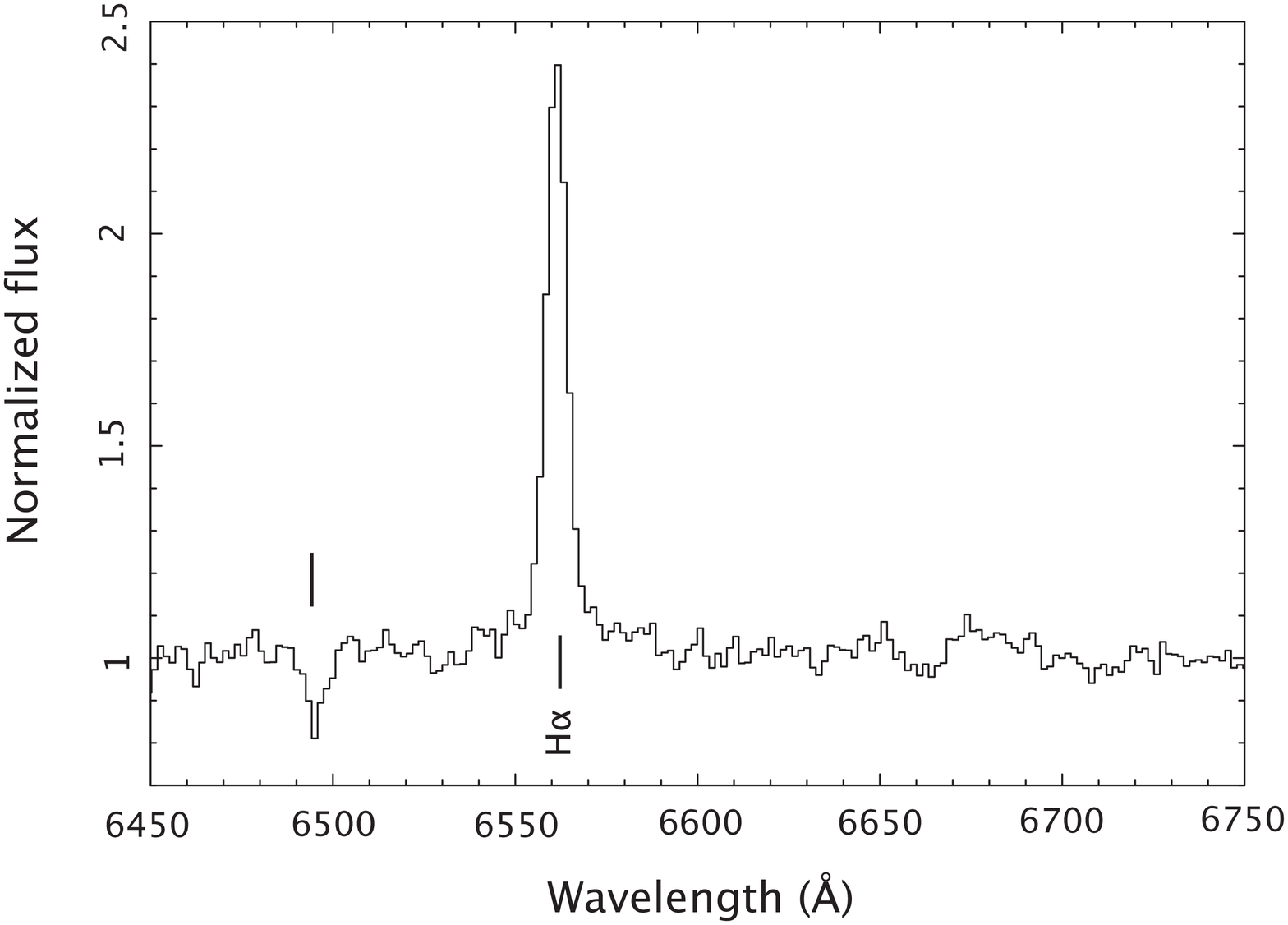}\hspace{0.1cm}
  	\includegraphics[width=8.0cm]{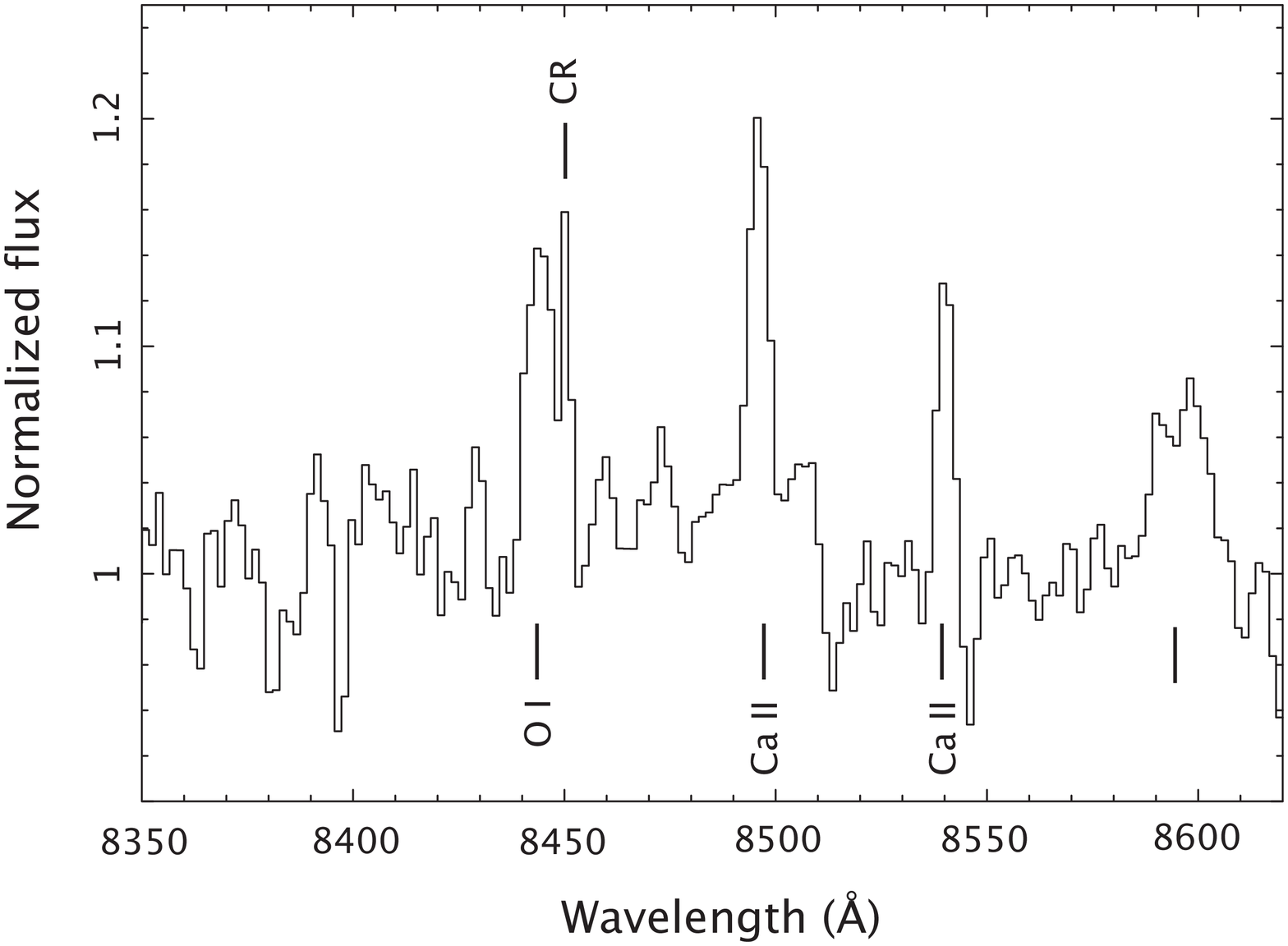}
    \end{center}
\caption[]{{Close-ups of the averaged VLT spectra. Top: region around \halfa. Bottom: region around the Ca\,{\sc ii} triplet. The feature labelled `CR' denotes a likely cosmic ray event.}}
 \label{fig:vltzoom}
\end{figure}

The \vlt\ spectra reveal several features, the most prominent being \halfa\ in emission above the continuum (see Fig.~\ref{fig:vltzoom}). A single Gaussian fit is a good representation of the line profile, yielding a full width at half maximum (FWHM) of $292 \pm
9~\kms$. The fits show that the line profile is blue-shifted $-58 \pm 4~\kms$ with respect to the rest wavelength, and no Doppler shifts in the central wavelength are seen between the three \vlt\ spectra (which were obtained in an interval of one hour). The line has an equivalent width (EW) of $-9.0\pm0.4$~\AA. 

The bottom plot of Fig.~\ref{fig:vltzoom} displays the region around the Ca\,{\sc ii} triplet (the redder component of the triplet falls outside the range covered by the detector). From a single Gaussian fit to the Ca\,{\sc ii} lines at 8498.02 and 8542.09~\AA, we find a blue shift of $-67 \pm 12~\kms$ and a FWHM=$276 \pm 12~\kms$. In addition, we detect O\,{\sc i} at 8446~\AA. The narrow feature next to it at $\sim8450$~\AA\ is likely a cosmic ray as it appears in only one of the three spectra. The feature at $\sim8590$~\AA\ could possibly be P\,{\sc 14} emission, although it seems to be too broad and the central wavelength does not agree with the shift observed for the other lines. The identification of this feature is therefore uncertain.

The main interstellar features are the sodium doublet at 5890 and 5896~\AA\ and interstellar bands at 5780 and 6284~\AA. Furthermore, an absorption is observed at 6495~\AA\ with an EW~= ~$1.5 \pm 0.2$ \AA\ (see Fig.~\ref{fig:vltzoom}). This feature is observed in late-type stars and is due to a blend of metallic absorption lines. However, we do not expect to observe photospheric lines in the spectrum of \rxh, as the accretion disc is likely to dominate the optical flux -- no other strong metallic lines are observed in the spectrum. A similar feature has been reported in the optical spectrum of the X-ray transient XTE~J1118+480 during outburst, which was tentatively associated to cool emitting regions in the accretion flow \citep[see][]{torres2002}.\\

\section{Discussion}\label{sec:discuss}
In this paper we presented X-ray, optical and near-IR observations of the previously unclassified \rosat\ source \rxh. This system was the likely origin of a \swift/BAT trigger that occurred on 2008 May 14 and we carried out a multi-wavelength observing 
campaign to get a more complete picture of the properties of this X-ray source.

\subsection{The optical/near-IR counterpart}\label{subsec:counterpart}
We investigated all the XRT and UVOT data that were obtained within two hours after the BAT trigger. The XRT spectra could be successfully modelled by blackbody radiation and revealed cooling during the decay, which confirms that this was a thermonuclear event. This testifies that \rxh\ contains an accreting neutron star and classifies the system, in all likelihood, as an LMXB. The UVOT $WH$-band images revealed an optical source that was fading simultaneously with the observed decrease in X-ray flux. Such behavior is typical of type-I X-ray bursts and is thought to result from reprocessing of X-rays \citep[e.g.,][]{lewin95}. This provides strong evidence that the fading UVOT source is the counterpart of \rxh. A similar fading was detected in $R$-band images obtained with the \rem\ telescope.

Using the \ntt, \magellan\ and \vlt, we detect an optical/near-IR source within the UVOT positional uncertainty of \rxh. The \vlt\ observations reveal a spectrum with a single-peaked \halfa\ emission line. Such emission is typical for X-ray binaries, accreting white dwarfs and Be stars. The broadband colours of the counterpart after correcting for the reddening are not consistent with a Be star, which has a bluer spectral energy distribution (SED) than observed \citep[cf. Section~\ref{subsec:photometry} and][]{drilling2000,tokunaga2000}. This effectively rules out the possibility that we detect a Be star interloper within the UVOT error circle. Thus, we conclude that we have detected the optical/near-IR counterpart of \rxh.

The \halfa\ and Ca\,{\sc ii} emission line broadening observed in the \vlt\ spectra is strongly affected by the instrumental profile, which makes it difficult to assess whether or not the lines are double-peaked. We subtract in quadrature the instrumental width to find a FWHM of $192 \pm 9$ and $225 \pm 12~\kms$ for the \halfa\ and Ca {\sc ii} lines respectively (see Section~\ref{subsec:specres}). The ratio of these FWHMs are consistent with the ratio of the rest wavelengths and thereby with Doppler broadening of the line.

The observed EW and intrinsic FWHM of the lines match two possible scenarios for the origin of the line emission. The first is that the emission arises from the accretion disc, in which case the line profile would be double-peaked unless if the system is viewed face-on \citep[e.g.,][]{huang72}. In the second scenario, the emission is due to X-ray reprocessing in the hemisphere of the secondary facing the neutron star, which would produce a single-peaked profile \citep[e.g.,][]{bassa09}. Further spectroscopic observations at higher spectral resolution may test these hypothesis.

\subsection{The type-I X-ray burst}\label{subsec:burst}
The parameters of the X-ray burst from \rxh, as inferred from spectral analysis of the BAT and XRT data, are summarized in Table~\ref{tab:burstpar}. These show that it was no ordinary type-I X-ray burst, which are triggered by unstable burning of H/He and typically last $\sim10-100$~s releasing a total energy of $\sim 10^{39-40}$~erg. Yet it is not as energetic as the so-called superbursts, which endure for many hours and are thought to be fuelled by carbon rather then H/He, resulting in a total energy release of $\sim 10^{42-43}$~erg \citep[e.g.,][]{strohmayer06}. Instead, the duration ($\sim 2$~h) and total energy output ($E_{\mathrm{burst}} \lesssim 1.5 \times10^{41}$~erg) suggest that the X-ray burst from \rxh\ belongs to the rare class of intermediately long X-ray bursts. The driving mechanism behind these events is thought to be the ignition of a thick layer of He and their host systems probe unusual accretion regimes \citep{zand05,zand07,cumming06}. 

Several intermediately long X-ray bursts have been detected from (candidate) ultra-compact X-ray binaries \citep[UCXBs; see e.g.,][]{zand05,zand08,falanga08,kuulkers09}. These systems have orbital periods $\lesssim 80$~min, which implies that the donor star must be H-deficient \citep[][]{nelson86}. In this context, the intermediately long bursts are explained in terms of the neutron stars accreting He-rich material. However, in case of \rxh\ the detection of strong \halfa\ emission in the spectrum of the optical counterpart strongly indicates that the donor star in \rxh\ is H-rich and that the neutron star is not accreting pure He.

There are a few other systems displaying intermediately long type-I X-ray bursts for which there are indications that the accreted matter contains H \citep[][]{chenevez07,linares09,falanga09}. \citet{peng2007} study the accretion of H-rich material at low accretion rates, and demonstrate that there exists a narrow regime, spanning only a factor of $\sim3$ in mass-accretion rate, for which H flashes occur that are too weak to ignite He. For accretion rates lower than this range, the rise in temperature following a H flash is sufficient to cause He ignition, resulting in a short, mixed H/He burst \citep{peng2007}. These authors speculate that a series of weak H flashes might build up a large reservoir of He that produces a long X-ray burst, like the ones observed for UCXBs, once it ignites. This behavior is confirmed by the theoretical models of \citet[][]{cooper07} and might provide the framework to explain the intermediately long X-ray burst observed from \rxh.

\citet{peng2007} find that unstable H burning can accumulate a thick layer of He for a regime of local mass-accretion rates of 0.3--1 per cent of the Eddington rate, which is higher than what we infer for \rxh\ ($\dot{m}\lesssim0.1$ per cent of Eddington; see Section~\ref{subsec:burst_ana}). However, the boundary values of this narrow range are sensitive to the heat flux emerging from the neutron star crust, which results from a series of non-equilibrium reactions induced by the accretion of matter \citep[see e.g.,][and references therein]{heansel08}. At low accretion rates, this heat flow largely sets the thermal structure of the accreted layer and thereby the ignition conditions for X-ray bursts.

To explain the occurrence of intermediately long X-ray bursts from systems accreting around 1 per cent of Eddington, \citet{peng2007} choose a heat release of 0.1~MeV per accreted nucleon. If this value is increased to 1.0~MeV, as may be better justified for the low accretion rates under consideration \citep{brown2004}, the range allowing for intermediately long bursts decreases to $\dot{m}\sim0.03-0.1$ per cent of Eddington, i.e., consistent with the value we infer for \rxh. Nevertheless, for this combination of heat release and $\dot{m}$, the expected ignition column depth is much higher than observed for \rxh\ \citep[see fig. 11 of][]{peng2007}.  

Achieving ignition at $y\lesssim1.5\times10^{10}~\mathrm{g~cm}^{-2}$, requires either that the heat deposited in the crust is more than 1.0~MeV per accreted nucleon \citep[which may be reasonable, see e.g.,][]{heansel08}, or that the local mass-accretion rate is actually higher than what we infer for \rxh. While we assumed isotropy, it is also possible that the accretion flow is concentrated onto a small area of the neutron star surface, in which case the local mass-accretion rate is underestimated. However, the apparent mismatch between the observations of \rxh\ and the theoretical calculations might also be due to limitations of the simplified model description \citep{peng2007}.

We note that the properties of the X-ray burst of \rxh\ are very similar to the intermediately long bursts from \xte\ \citep[][]{linares09,falanga09} and \ucxb\ \citep[][]{zand08}, which both triggered the BAT and were subsequently observed by the XRT. The three X-ray bursts have similar BAT rise times of tens of seconds and we found that the XRT tails show comparable decay rates. Yet the nature of the three systems seems to be very different. 
\xte\ is known to be a transient system, albeit it is exhibiting a prolonged accretion outburst that started in 2008 June and is ongoing at the time of writing \citep[the intermediately long burst was detected 5 weeks after the onset of the outburst; e.g.,][]{linares09}. \ucxb, on the other hand, is persistently accreting and is proposed to be an UCXB based on its low optical magnitude and the absence of \halfa\ emission in the source spectrum \citep{bassa06}. Both systems accreted at a level of $\sim$1 per cent of  the Eddington rate when the intermediately long X-ray bursts occurred. \rxh\ seems to be a persistent system (see Section~\ref{subsec:nature}) accreting from a H-rich donor at $\sim$0.1 per cent of  the Eddington rate, which is a factor of 10 lower than inferred for the other two sources.

\subsection{The nature of \rxhname}\label{subsec:nature}
\rxh\ was detected with \rosat\ in 1990 and 1994, with \xmm\ in 2008 March, and the source field was covered several times with \swift\ between 2008 May--August, as well as during a single pointing in 2009 July. On all occasions the source displayed similar unabsorbed fluxes of $\sim (1-4) \times 10^{-11}~\flux$ (0.5--10 keV), which indicates that \rxh\ is a persistent, rather than transient, LMXB. The long burst recurrence time confirms that the system is intrinsically faint and accreting at low rates. The persistent nature at a low accretion luminosity suggests the possibility of a relatively small orbit. Small accretion discs are easier to be kept photo-ionized completely, thereby sustaining the accretion and avoiding the disc instability model that would make the system transient. Based on this argument, \citet{zand07} use a low persistent flux as a diagnostic to put forward several candidate UCXBs, drawn from the total sample of bursting, persistent LMXBs. However, our optical data suggests that \rxh\ is likely not an UCXB.

As already mentioned above, the detection of strong \halfa\ emission in the spectrum of the optical counterpart strongly indicates that the donor star in \rxh\ is H-rich, effectively ruling out the UCXB scenario. Furthermore, the absolute visual magnitude ($M_V$) of \rxh\ can be estimated using the distance modulus. For a de-reddened apparent magnitude of $V=16.0$ and a distance $D \lesssim 9.5$~kpc (inferred from the peak of the X-ray burst), we find an absolute visual magnitude of $M_V \gtrsim 1.1$~mag. For the estimated mass-accretion rate of \rxh\ (0.1 per cent of Eddington, see Section~\ref{subsec:burst_ana}), the empirical relation derived by \citet{vanparadijs94} predicts an absolute visual magnitude of $M_V \gtrsim 4.8$~mag in case the system is an UCXB (assuming $P_{\mathrm{orb}} \lesssim 80$~min). Unless \rxh\ is located at a distance $D \lesssim 2$~kpc, it is thus too optically bright to be an UCXB.  

To be able to harbour a H-rich companion, \rxh\ must have an orbital period of $\gtrsim 80$~min \citep[e.g.,][]{nelson86}. In such a configuration, it will be challenging to understand how the low X-ray luminosity can keep the accretion ongoing making the system persistent rather than transient.

\begin{table}
\begin{center}
\caption[l]{X-ray burst and persistent emission parameters.}
\begin{threeparttable}
\begin{tabular}{l l}
\hline \hline
\textbf{X-ray burst} & \\
BAT rise time (s) & $\sim 100$ \\
Duration (h) & $\sim 2.2$ \\
Peak flux, $F_{\mathrm{bol}}^{\mathrm{peak}}$ ($\flux$) & $\sim3.5\times 10^{-8}$ \\
Fluence, $f_{\mathrm{burst}}$ ($\mathrm{erg~cm}^{-2}$) & $\sim1.5 \times 10^{-5}$ \\
Distance, $D$ (kpc) & $\lesssim 9.5$ \\
Total radiated energy, $E_{\mathrm{burst}}$ (erg) & $\lesssim 1.6 \times 10^{41}$ \\
Ignition depth, $y$ ($\mathrm{g~cm}^{-2}$) & $\lesssim 1.5 \times 10^{10}$ \\
& \\ 
\hline
\textbf{Persistent emission} & \\
Flux, $F_{\mathrm{bol}}^{\mathrm{pers}}$ ($\flux$) & $\sim3.8 \times 10^{-11}$\\
Luminosity, $L_{\mathrm{bol}}^{\mathrm{pers}}$ ($\lum$) & $\lesssim4.1 \times 10^{35}$\\
Global accretion rate, $\dot{M}$ ($\Msun~\mathrm{yr}^{-1}$) & $\lesssim3.6 \times 10^{-11}$\\
Local accretion rate, $\dot{m}$ ($\mathrm{g~cm}^{-2}~\mathrm{s}^{-1}$) & $\lesssim1.7 \times 10^{2}$\\
\hline
\end{tabular}
\label{tab:burstpar}
\begin{tablenotes}
\item[] \textit{Note}. The quoted fluxes are unabsorbed and for the 0.01--100 keV energy range. The burst duration is specified as the time from the BAT peak till the flux decayed back to the persistent level as observed with the XRT.
\end{tablenotes}
\end{threeparttable}
\end{center}
\end{table}

\section*{Acknowledgments}
We are grateful to the referee, Craig Heinke, for useful comments that helped improve this manuscript. 
This work was based on observations made with ESO Telescopes at the Paranal and La Silla Observatories under programme IDs: 281.D-5030(A) and 60.A-9700(D) and made use of the public data archive of \swift\ and \inte, as well as public data from the \xmm\ slew survey. 
Support for this work was provided by the Netherlands Organization for Scientific Research (NWO).  NR acknowledges support from a Ramon y Cajal Research position. EMC gratefully acknowledges support provided by NASA through the \chan\ Fellowships Program, grant number PF8-90052. We acknowledge the use of the software package \textsc{molly} written by Prof.~Tom Marsh.

\bibliographystyle{mn2e}

\label{lastpage}
\end{document}